\documentclass[conference]{src/IEEEtran}
\IEEEoverridecommandlockouts

\usepackage{amsmath,amssymb,amsfonts,amsthm,pifont}
\usepackage{graphicx,import}
\usepackage{textcomp}
\usepackage{xcolor}
\def\BibTeX{{\rm B\kern-.05em{\sc i\kern-.025em b}\kern-.08em
		T\kern-.1667em\lower.7ex\hbox{E}\kern-.125emX}}

\usepackage[style=ieee, backend=bibtex, doi=false, url=false, isbn=false]{biblatex} 

\usepackage{algorithm}
\usepackage{algpseudocode} 
\usepackage{hyperref}
\hypersetup{hidelinks}
\usepackage{textcomp}
\usepackage[nolist, nohyperlinks, printonlyused, withpage, smaller]{acronym}
\usepackage{booktabs,bigstrut,rotating}  

\usepackage{tablefootnote}

\usepackage{colortbl}  

\usepackage{enumitem}
\usepackage{tikz}
\usetikzlibrary{tikzmark, calc}
\usetikzlibrary{positioning, calc}

\newcommand{\cmark}{\textcolor[RGB]{0,120,0}{\scalebox{1.2}{\ding{51}}}}%
\newcommand{\xmark}{\textcolor[RGB]{120,0,0}{\scalebox{1.2}{\ding{55}}}}%

\newcommand{\rot}[1]{\begin{sideways}\shortstack[l]{#1}\end{sideways}}

\newcommand\copyrighttext{%
	\footnotesize Contribution to the 29th International Conference on Information Fusion. \copyright 2026 IEEE.  Personal use of this material is permitted. Permission from IEEE must be obtained for all other uses, in any current or future media, including reprinting/republishing this material for advertising or promotional purposes, creating new collective works, for resale or redistribution to servers or lists, or reuse of any copyrighted component of this work in other works.}

\newcommand\copyrightnotice{%
	\begin{tikzpicture}[remember picture,overlay]
		\node[anchor=south,yshift=7mm] at (current page.south) {\fbox{\parbox{\dimexpr\textwidth-\fboxsep-\fboxrule\relax}{\copyrighttext}}};
	\end{tikzpicture}%
}

\begin{document}
	

\begin{acronym}
	\acro{rnn}[RNN]{Recurrent Neural Network}
	\acroplural{rnn}[RNN]{Recurrent Neural Networks}
	\acro{narx}[NARX-NN]{Nonlinear Autoregressive Exogenous Neural Network}
	\acroplural{narx}[NARX-NN]{Nonlinear Autoregressive Exogenous Neural Networks}
	\acro{gru}[GRU]{{Gated Recurrent Unit}}
	\acroplural{gru}[GRU]{{Gated Recurrent Units}}
	\acro{lstm}[LSTM]{Long Short-Term Memory}
	\acro{ann}[NN]{Neural Network}	
	\acroplural{ann}[NN]{Neural Networks}
	\acro{ffnn}[FFNN]{Feedforward Neural Network}
	\acroplural{ffnn}[FFNN]{Feedforward Neural Networks}
	\acro{pinn}[PINN]{Physics-Informed Neural Network}
	\acroplural{pinn}[PINN]{Physics-Informed Neural Networks}
	\acro{gp}[GP]{Gaussian Process}
	\acroplural{gp}[GPs]{Gaussian Processes}
	\acro{knn}[$K$NN]{$K$-Nearest-Neighbors}
	\acro{ilc}[ILC]{Iterative Learning Control}
	\acro{ili}[ILI]{Iterative Learning Identification}
	\acro{rls}[RLS]{Recursive Least Squares}
	\acro{rc}[RC]{Repetitive Control}
	\acro{rl}[RL]{{Reinforcement Learning}}
	\acro{daoc}[DAOC]{{Direct Adaptive Optimal Control}}
	\acro{ml}[ML]{maschinelles Lernen}
	\acro{lwpr}[LWPR]{{Locally Weighted Projection Regression}}
	\acro{svm}[SVM]{{Support Vector Machine}}
	\acro{mcmc}[MCMC]{{Markov Chain Monte Carlo}}
	\acro{ad}[AD]{{Automatic Differentiation}}
	\acro{gmm}[GMM]{{Gaussian Mixture Models}}
	\acro{rkhs}[RKHS]{{Reproducing Kernel Hilbert Spaces}}
	\acro{rbf}[RBF]{{Radial Basis Function}}
	\acro{rbfnn}[RBF-NN]{{Radial Basis Function Neural Network}}
	\acroplural{rbfnn}[RBF-NN]{{Radial Basis Function Neural Networks}}
	\acro{node}[{Neural} ODE]{{Neural Ordinary Differential Equation}}
	\acroplural{node}[{Neural} ODEs]{{Neural Ordinary Differential Equations}}
	\acro{pdf}[PDF]{Probability Density Function}
	\acro{pca}[PCA]{Principal Component Analysis}
	\acro{lnn}[LNN]{Lagrangian Neural Network}
	\acroplural{lnn}[LNNs]{Lagrangian Neural Networks}
	\acro{hnn}[HNN]{Lagrangian Neural Network}
	
	\acroplural{hnn}[HNNs]{Hamiltonian Neural Networks}

	\acro{kf}[KF]{Kalman Filter}	
	\acro{ekf}[EKF]{Extended Kalman Filter}
	\acro{nekf}[NEKF]{Neural Extended Kalman Filter}
	\acro{ukf}[UKF]{Unscented Kalman Filter}
	\acro{pf}[PF]{Particle Filter}
    \acroplural{pf}[PFs]{Particle Filters}
	\acro{mhe}[MHE]{{Moving Horizon Estimation}}
	\acro{rpe}[RPE]{{Recursive Predictive Error}}
	\acro{slam}[SLAM]{{Simultaneous Location and Mapping}}
	
	\acro{mftm}[MFTM]{Magic Formula Tire Model}
	\acro{mbs}[MBS]{Multi-Body Simulation}
	\acro{lti}[LTI]{Linear Time-Invariant}
	\acro{cog}[COG]{Center Of Gravity}
	\acro{ltv}[LTV]{Linear Time-Variant}
	\acro{siso}[SISO]{Single Input Single Output}
	\acro{mimo}[MIMO]{Multiple Input Multiple Output}
	\acro{psd}[PSD]{Power Spectral Density}
	\acroplural{psd}[PSD]{Power Spectral Densities}
	\acro{cf}[CF]{Coordinate Frame}
	\acroplural{cf}[CF]{Coordinate Frames}
	\acro{phs}[PHS]{Port-Hamiltonian System}

	\acro{pso}[PSO]{Particle Swarm Optimization}
	\acro{sqp}[SQP]{Sequentielle Quadratische Programmierung}
	\acro{svd}[SVD]{Singular Value Decomposition}
	\acro{ode}[ODE]{Ordinary Differential Equation}
	\acroplural{ode}[ODE]{Ordinary Differential Equations}
	\acro{pde}[PDE]{Partial Differential Equation}
	
	\acro{nmse}[NMSE]{Normalized Mean Squared Error}
	\acro{mse}[MSE]{Mean Squared Error}
	\acro{rmse}[RMSE]{Root Mean Squared Error}
	\acro{wrmse}[wRMSE]{weighted \ac{rmse}}
	\acro{nrmse}[NRMSE]{Normalized Root Mean Squared Error}
	
	\acro{mpc}[MPC]{{Model Predictive Control}}
	\acro{nmpc}[NMPC]{Nonlinear Model Predictive Control}
	\acro{lmpc}[LMPC]{Learning Model Predictive Control}
	\acro{ltvmpc}[LTV-MPC]{\acl{ltv} Model Predictive Control}
	\acro{ndi}[NDI]{Nonlinear Dynamic Inversion}
	\acro{ac}[AC]{Adhesion Control}
	\acro{esc}[ESC]{Electronic Stability Control}
	\acro{ass}[ASS]{Active Suspension System}
	\acro{trc}[TRC]{Traction Control}
	\acro{abs}[ABS]{Anti-Lock Brake System}
	\acro{gcc}[GCC]{Global Chassis Control}
	\acro{ebs}[EBS]{Electronic Braking System}
	\acro{adas}[ADAS]{Advanced Driver Assistance Systems}
	\acro{cm}[CM]{{Condition Monitoring}}
	\acro{hil}[HiL]{{Hardware-in-the-Loop}}
	\acro{siso}[SISO]{Single Input Single Output}
	\acro{mimo}[MIMO]{Multiple Input Multiple Output}

	\acro{irw}[IRW]{Independently Rotating Wheels}
	\acro{dirw}[DIRW]{Driven \acl{irw}}
	\acro{imes}[imes]{{Institute of Mechatronic Systems}}
	\acro{db}[DB]{Deutsche Bahn}
	\acro{ice}[ICE]{Intercity-Express}
	
	\acro{fmi}[FMI]{{Functional Mock-up Interface}}
	\acro{fmu}[FMU]{{Functional Mock-up Unit}}
	\acro{doi}[DOI]{{Digital Object Identifier}}

	\acro{ra}[RA]{Research Area}
	\acroplural{ra}[RA]{Research Areas}
	\acro{wp}[WP]{Work Package}
	\acroplural{wp}[WP]{Work Packages}
	
	\acro{fb}[FB]{Forschungsbereich}
	\acroplural{fb}[FB]{Forschungsbereiche}
	\acro{ap}[AP]{Arbeitspaket}
	\acroplural{ap}[AP]{Arbeitspakete}
	\acro{abb}[Abb.]{Abbildung}
	\acro{luis}[LUIS]{Leibniz Universit"at IT Services}
	
	\acro{res}[RES]{Renewable Energy Sources}
	\acro{pkw}[PKW]{Personenkraftwagen}
	\acro{twipr}[TWIPR]{{Three-Wheeled Inverted Pendulum Robot}}
	\acro{tlr}[TLR]{Two-Link Robot}
	
	\acro{pmcmc}[PMCMC]{Particle Markov Chain Monte Carlo}
	\acro{mcmc}[MCMC]{Markov Chain Monte Carlo}
	\acro{rbpf}[RBPF]{Rao-Blackwellized Particle Filter}
	\acro{pf}[PF]{Particle Filter}
	\acro{ps}[PS]{Particle Smoother}
	\acro{smc}[SMC]{Sequential Monte Carlo}
	\acro{csmc}[cSMC]{conditional SMC}
	\acro{mh}[MH]{Metropolis Hastings}
	\acro{em}[EM]{Expectation Maximization}
	\acro{slam}[SLAM]{Simultaneous Location and Mapping}
	\acro{dof}[DOF]{Degree of Freedom}
	\acroplural{dof}[DOF]{Degrees of Freedom}
	\acro{pg}[PG]{Particle Gibbs}
	\acro{pgas}[PGAS]{Particle Gibbs with Ancestor Sampling}
	\acro{mpgas}[mPGAS]{marginalized Particle Gibbs with Ancestor Sampling}
	\acro{hmm}[HMM]{Hidden Markov Model}
	
	\acro{emps}[EMPS]{Electro-Mechanical Positioning System}
	\acro{fts}[FTS]{Force-Torque Sensor}

	\acro{ilc}[ILC]{Iterative Learning Control}
	\acro{ddilc}[DD-ILC]{Data-Driven Iterative Learning Control}
	\acro{dilc}[DILC]{Dual Iterative Learning Control}
	\acro{iml}[IML]{Iterative Model Learning}
	\acro{noilc}[NO-ILC]{Norm-Optimal Iterative Learning Control}
	\acro{gilc}[G-ILC]{Gradient Iterative Learning Control}

	\acro{bilbo}[BILBO]{Balancing Intelligent Learning roBOt}

\end{acronym}

\newcommand{\qvec}[1]{\mathbf{#1}}
\newcommand{\qmat}[1]{\mathbf{#1}}
\newcommand{\idx}[1]{_{\mathrm{#1}}}

\newcommand{\qRealNumbers}{\mathbb{R}}
\newcommand{\qPositiveRealNumbers}{\mathbb{R}_{\geq 0}}
\newcommand{\qNaturalNumbersZero}{\mathbb{N}_{\geq 0}}
\newcommand{\qNaturalNumbersPos}{\mathbb{N}_{>0}}
\newcommand{\qNaturalNumbers}{\mathbb{N}}
\newcommand{\foralljinN}{\forall j \in \qNaturalNumbersZero, \quad}
\newcommand{\foralljgeqN}{\forall j \in \qNaturalNumbersPos, \quad}
\newcommand{\qBLT}{\mathcal{T}^{\mathrm{BLT}}}
\newcommand{\qNormal}{\mathcal{N}}

\newcommand\invChi{\mathop{\mbox{Scale-inv-$\chi^2$}}}

\newcommand{\qa}{\qvec{a}}
\newcommand{\qb}{\qvec{b}}
\newcommand{\qd}{\qvec{d}}
\newcommand{\qe}{\qvec{e}}
\newcommand{\qf}{\qvec{f}}
\newcommand{\qg}{\qvec{g}}
\newcommand{\qh}{\qvec{h}}
\newcommand{\qi}{\qvec{i}}
\newcommand{\qj}{\qvec{j}}
\newcommand{\qk}{\qvec{k}}
\newcommand{\ql}{\qvec{l}}
\newcommand{\qm}{\qvec{m}}
\newcommand{\qn}{\qvec{n}}
\newcommand{\qo}{\qvec{o}}
\newcommand{\qp}{\qvec{p}}
\newcommand{\qr}{\qvec{r}}
\newcommand{\qs}{\qvec{s}}
\newcommand{\qt}{\qvec{t}}
\newcommand{\qu}{\qvec{u}}
\newcommand{\qv}{\qvec{v}}
\newcommand{\qw}{\qvec{w}}
\newcommand{\qx}{\qvec{x}}
\newcommand{\qy}{\qvec{y}}
\newcommand{\qz}{\qvec{z}}
\newcommand{\qem}{\qvec{e}^{\mathrm{m}}}

\newcommand{\qA}{\qvec{A}}
\newcommand{\qB}{\qvec{B}}
\newcommand{\qC}{\qvec{C}}
\newcommand{\qD}{\qvec{D}}
\newcommand{\qE}{\qvec{E}}
\newcommand{\qF}{\qvec{F}}
\newcommand{\qG}{\qvec{G}}
\newcommand{\qH}{\qvec{H}}
\newcommand{\qI}{\qvec{I}}
\newcommand{\qJ}{\qvec{J}}
\newcommand{\qK}{\qvec{K}}
\newcommand{\qL}{\qvec{L}}
\newcommand{\qM}{\qvec{M}}
\newcommand{\qN}{\qvec{N}}
\newcommand{\qO}{\qvec{O}}
\newcommand{\qP}{\qvec{P}}
\newcommand{\qQ}{\qvec{Q}}
\newcommand{\qR}{\qvec{R}}
\newcommand{\qS}{\qvec{S}}
\newcommand{\qT}{\qvec{T}}
\newcommand{\qU}{\qvec{U}}
\newcommand{\qV}{\qvec{V}}
\newcommand{\qW}{\qvec{W}}
\newcommand{\qX}{\qvec{X}}
\newcommand{\qY}{\qvec{Y}}
\newcommand{\qZ}{\qvec{Z}}

\newcommand{\quv}{\bar{\qu}}
\newcommand{\qyv}{\bar{\qy}}
\newcommand{\qxv}{\bar{\qx}}
\newcommand{\qZero}{\qvec{0}}
\newcommand{\qdu}{\boldsymbol{\Delta}\qu}
\newcommand{\qdy}{\boldsymbol{\Delta}\qy}
\newcommand{\qep}{\hat{\qe}}
\newcommand{\qyp}{\hat{\qy}}

\newcommand{\qILCDesign}{\boldsymbol{D}}
\newcommand{\qIMLDesign}{\boldsymbol{\hat{D}}}
\newcommand{\qLIML}{\hat{\qL}}
\newcommand{\qeR}{\qe_\mathrm{R}}


\newcommand{\qff}{\boldsymbol{f}}
\newcommand{\qpf}{\boldsymbol{p}}
\newcommand{\qmf}{\boldsymbol{m}}
\newcommand{\qXf}{\boldsymbol{A}}

\newcommand{\qLift}{\mathcal{L}}
\newcommand{\qLu}{\boldsymbol{\mathcal{L}\idx{u}}}
\newcommand{\qLm}{\boldsymbol{\mathcal{L}\idx{m}}}
\newcommand{\qUb}{\bar{\qU}}
\newcommand{\qyh}{\hat{\qy}}
\newcommand{\qeh}{\hat{\qe}}
\newcommand{\qLh}{\hat{\qL}}
\newcommand{\qDeh}{\hat{\boldsymbol{\mathcal{D}}}}
\newcommand{\qDe}{{\boldsymbol{\mathcal{D}}}}
\newcommand{\qMt}{\tilde{\qM}}
\newcommand{\qPt}{\tilde{\qP}}
\newcommand{\qLt}{\tilde{\qL}}
\newcommand{\qQt}{\tilde{\qQ}}
\newcommand{\qSt}{\tilde{\qS}}
\newcommand{\qWt}{\tilde{\qW}}
\newcommand{\qRt}{\tilde{\qR}}
\newcommand{\qut}{\tilde{\qu}}
\newcommand{\qyt}{\tilde{\qy}}
\newcommand{\qet}{\tilde{\qe}}
\newcommand{\qzt}{\tilde{\qz}}
\newcommand{\qrt}{\tilde{\qr}}
\newcommand{\qXt}{\tilde{\qX}}

\newcommand{\qTb}{\bar{\qT}}
\newcommand{\qVb}{\bar{\qV}}

\newcommand{\qQh}{\hat{\qQ}}
\newcommand{\qSh}{\hat{\qS}}
\newcommand{\qXh}{\hat{\qX}}
\newcommand{\qWh}{\hat{\qW}}

\newcommand{\Lfunc}{\mathscr{L}}       
\newcommand{\Kfunc}{\mathscr{K}}
\newcommand{\KLfunc}{\Kfunc\negthinspace\negthinspace\Lfunc}

\newcommand{\qTwoNorm}[1]{\left\| {#1} \right\|_{2}}
\newcommand{\qInfNorm}[1]{\left\| {#1} \right\|_\infty}
\newcommand{\qOneNorm}[1]{\left\| {#1} \right\|\idx{1}}
\newcommand{\qNorm}[1]{\left\| {#1} \right\|}
\newcommand{\qGivenNorm}{\qNorm{\boldsymbol{\cdot}}}
\newcommand{\norm}[1]{\left\| {#1} \right\|}

\newcommand{\qred}[1]{{\color{red}#1}}
\newcommand{\imesorange}{E77B29}
\newcommand{\imesgruen}{C8D317}
\newcommand{\imesblauHundert}{00509B}
\newcommand{\imesblauZwanzig}{CCDCEB}
\newcommand{\imesblauVierzig}{99B9D8}

\newcommand{\eg}{e.\,g.,\,}
\newcommand{\ie}{i.\,e.,\,}
\newcommand*{\R}{\mathbb{R}}

\newcommand*{\MNIW}{\mathcal{MNIW}}
\newcommand*{\IW}{\mathcal{IW}}

\newcommand{\del}{\partial}
\newcommand{\bi}[1]{\boldsymbol{#1}}
\newcommand{\ur}[1]{\mathrm{#1}}
\newcommand{\cali}[1]{\mathcal{#1}}

\newcommand{\ubar}[1]{\underaccent{\bar}{#1}}
\newcommand{\ToDo}[1]{\todo[size=\tiny]{#1}}

\newcommand{\spans}[1]{\spanop\left( #1 \right)}

\newcommand{\ToDos}[1]{\todo[size=\tiny]{#1}}


	\title{Simultaneous State Estimation and \\ Online Model Learning in a Soft Robotic System \thanks{This research was supported in parts by the \textit{Kjell och Märta Beijer Foundation} and the \textit{Swedish Research Council (VR)} under the contract numbers 2021-04321 and 2025-04318, and the \textit{German Academic Scholarship Foundation (Studienstiftung des Deutschen Volkes)}.}
	}
	
	
		\author{\IEEEauthorblockN{Jan-Hendrik Ewering$^{\mathrm{a,b}}$}
		\IEEEauthorblockA{\hspace{0.3cm}\href{mailto:ewering@imes.uni-hannover.de}{ewering@imes.uni-hannover.de}\hspace{0.3cm}}
		\and
		\IEEEauthorblockN{Max Bartholdt$^{\mathrm{a,c}}$}
		\IEEEauthorblockA{\hspace{0.3cm}\href{mailto:bartholdt.max@mh-hannover.de}{bartholdt.max@mh-hannover.de}\hspace{0.3cm}}
		\and
		\IEEEauthorblockN{Simon F. G. Ehlers$^{\mathrm{a}}$}
		\IEEEauthorblockA{\hspace{0.3cm}\href{mailto:ehlers@imes.uni-hannover.de}{ehlers@imes.uni-hannover.de}\hspace{0.3cm}}
		\and
		\IEEEauthorblockN{Niklas Wahlstr{\"o}m$^{\mathrm{b}}$}
		\IEEEauthorblockA{\hspace{0.3cm}\href{mailto:niklas.wahlstrom@it.uu.se}{niklas.wahlstrom@it.uu.se}\hspace{0.3cm}}
		\and
		\IEEEauthorblockN{Thomas B. Sch{\"o}n$^{\mathrm{b}}$}
		\IEEEauthorblockA{\hspace{0.3cm}\href{mailto:thomas.schon@uu.se}{thomas.schon@uu.se}\hspace{0.3cm}}
		\and
		\IEEEauthorblockN{Thomas Seel$^{\mathrm{a}}$}
		\IEEEauthorblockA{\hspace{0.3cm}\href{mailto:seel@imes.uni-hannover.de}{seel@imes.uni-hannover.de}\hspace{0.3cm}}
		\and
		\IEEEauthorblockA{$^{\mathrm{a}}$\textit{Institute of Mechatronic Systems}, \textit{Leibniz Universit{\"a}t Hannover},	Hanover, Germany. \hfill}
		\IEEEauthorblockA{$^{\mathrm{b}}$\textit{Department of Information Technology}, \textit{Uppsala University},	Uppsala, Sweden. \hfill}
		\IEEEauthorblockA{$^{\mathrm{c}}$\textit{Clinical Dep. of Cardiothoracic, Transplantation, and Vascular Surgery}, \textit{Hannover Medical School}, Hanover, Germany. \hfill}
	}
    
	\maketitle
    \copyrightnotice\vspace{-9.5pt}
	
	\begin{abstract}
		Operating complex real-world systems, such as soft robots, can benefit from precise predictive control schemes that require accurate state and model knowledge. 
		This knowledge is typically not available in practical settings and must be inferred from noisy measurements. 
        In particular, it is challenging to simultaneously estimate unknown states and learn a model online from sequentially arriving measurements. 
		In this paper, we show how a recently proposed gray-box system identification tool enables the estimation of a soft robot's current pose while at the same time learning a bending stiffness model. 
        For estimation and learning, we only need a \emph{nominal} constant-curvature robot model and measurements of the robot's base reactions (\eg base forces). 
        The estimation scheme---relying on a marginalized particle filter---allows us to conveniently interface nominal constant-curvature equations with a \ac{gp} bending stiffness model to be learned. 
        This, in contrast to estimation via a random walk over stiffness values, enables prediction of bending stiffness and improves overall model quality. 
		We demonstrate, using a real-world soft robot, that the method learns a bending-stiffness model online while accurately estimating the robot's pose. 
		Notably, reduced error in multi-step forward predictions indicates that the learned bending-stiffness \ac{gp} improves overall model quality. 
	\end{abstract}
	
	
	\section{Introduction}
    Predictive control schemes, such as \ac{mpc}, require system models that provide accurate multi-step forward predictions \cite{Licher.2025,Hachen.2025,Vaskov.2023}. 
    Moreover, state information, \eg position and velocity data, is usually required to reliably operate complex real-world systems, such as autonomous vehicles \cite{Vaskov.2023} or soft robots \cite{Licher.2025,Hachen.2025}. 
    However, in many practical applications, only nominal models are available, and the unknown system state is just indirectly observed through output measurements. 
    In this case, we face a coupled inference problem: 
    Estimating system states from measurement data often relies on model knowledge \cite{Sarkka.2023}. 
    Conversely, model learning schemes often rely on state measurements. 
    Therefore, jointly learning a model and inferring unknown states is challenging, especially in online and nonlinear filtering applications. 


    \begin{figure}[tb] 
		\centering
		{\fontsize{7pt}{7pt}\selectfont
			\resizebox{0.85\linewidth}{!}{
				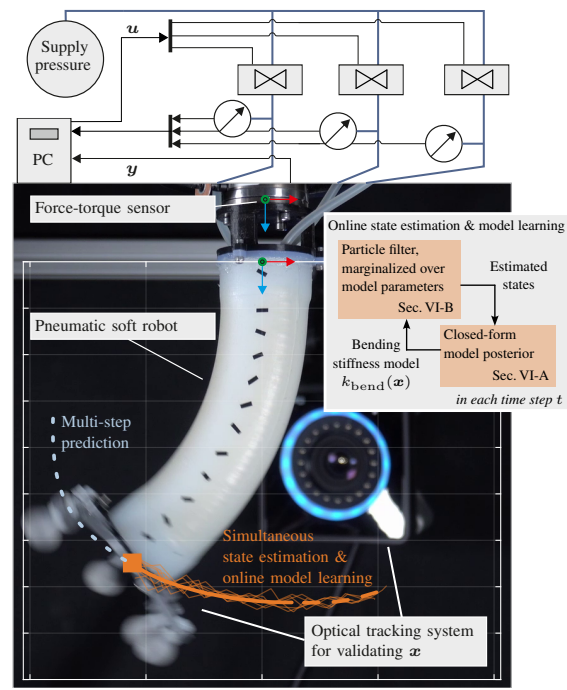
			}
		}
		\caption{Nonlinear soft robotic system with the unknown states $\boldsymbol{x} \in \mathbb{R}^{6}$, defining its current pose and velocity, control input pressures $\boldsymbol{u} \in \mathbb{R}^{3}$, and force/torque measurements of the base reactions $\boldsymbol{y} \in \mathbb{R}^{3}$. Simultaneously to estimating the current hidden states $\boldsymbol{x}$, a state-dependent bending stiffness model $k_{\mathrm{bend}} (\boldsymbol{x})$ is learned using a particle filter, marginalized over model parameters. Figure adapted from \cite{Mehl.2024} and reprinted with permission from 2024 IEEE Int. Conf. on Robotics and Automation (ICRA). \copyright 2024 IEEE.}
		\label{fig:problem_setting}
	\end{figure}

    In this paper, we face this joint inference problem in the soft robotic system depicted in Figure\,\ref{fig:problem_setting}. 
    This is interesting, as soft and continuum robots, often driven by pneumatic pressure \cite{Mehl.2024} or tendons \cite{Hachen.2025,Teetaert.2025b}, are compliant and flexible by construction \cite{DellaSantina.2023}, thereby paving the way for novel application domains. 
    However, it is often difficult to construct accurate dynamics models and to sense a soft robot's current pose \cite{DellaSantina.2023,Teetaert.2025}. 
    
    Specifically, we show how to estimate a soft robot's current pose while, for the first time, simultaneously learning a bending stiffness model online. 
    This is enabled by a recently proposed gray-box system identification tool \cite{Volkmann.2025}, which we validate on a complex, real-world robotic system, in contrast to the vast majority of work on joint state inference and model learning. 
    Unlike most soft robotic estimation schemes, we do not rely on restrictive sensor setups, such as optical tracking, but use only noisy measurements of the robot's base reactions (\eg base forces). 
    
    To this end, we employ the tailored \ac{pf} for gray-box system identification proposed in \cite{Volkmann.2025} and extend it with a hyperparameter learning procedure. 
    Importantly, the estimation scheme allows us to conveniently interface a nominal constant-curvature approximation with a \acl{gp} (\acs{gp}) bending stiffness model, which is to be learned. 
    This retains physical model interpretability by explicitly learning the bending stiffness model \textit{nested within} a well-known nonlinear constant-curvature model, differing from black-box learning approaches. 
    Moreover, we showcase in a multi-step prediction test scenario that the model quality is significantly improved through online learning.

    \section{Related Work}
    
    Considering the general methodology, there are various works on Bayesian state inference and model learning, often built around nonlinear system identification methods \cite{Wigren.2022}. 
    A natural extension for Bayesian modeling is to learn a \ac{gp} state-space model from batched \cite{Turner.2010,Frigola.2013,Svensson.2016,Svensson.2017,Ewering.2025} or---as done in this work---sequentially arriving input-output data \cite{Berntorp.2021,Berntorp.2022,Ewering.2024b,Volkmann.2025,Kullberg.2021c,Gotte.2023b}. 
    To enable model learning, many approximate the states as auxiliary variables using nonlinear Kalman filters \cite{Kullberg.2021c,Gotte.2023b} or \ac{smc} approaches \cite{Frigola.2013,Svensson.2016,Svensson.2017,Ewering.2025,Berntorp.2021,Berntorp.2022,Ewering.2024b,Volkmann.2025}. 
    The latter has the advantage that complex non-Gaussian probability density functions can be conveniently represented, which we exploit in this paper. 
    While most work focuses on learning an \textit{entire} \ac{gp} state-space model \cite{Berntorp.2021,Berntorp.2022,Gotte.2023b}, fewer methods enable incorporating system knowledge in the form of a nonlinear nominal model \cite{Ewering.2024b,Volkmann.2025,Kullberg.2021c}, as readily available for the current soft robotic application. 
    While all of the above research enables simultaneous state estimation and online model learning, none of the presented methods has been validated in a nonlinear real-world robotic system. 

	%
	%
	%
	%
	%
    
    In the subfield of soft robotic state estimation, existing work focuses either on offline estimation of the robot's shape \cite{Lilge.2022,Teetaert.2025} or on online estimation from sequentially arriving measurements \cite{Mehl.2024,Ataka.2016,Loo.2019,Lobaton.2013,Zhang.2026,Abdelaziz.2023,Kim.2021,Teetaert.2025b}. 
    For online applications, while some papers use sliding-window observers \cite{Abdelaziz.2023,Teetaert.2025b}, the predominant estimation approach is Bayesian filtering \cite{Mehl.2024,Ataka.2016,Loo.2019,Zhang.2026,Kim.2021}. 
    Importantly, all of these approaches rely on system models. 
    Therefore, estimation schemes build on Cosserat rod models \cite{Lilge.2022,Teetaert.2025,Teetaert.2025b} or, more commonly, on simplified nominal models, using constant-curvature assumptions or polynomial shape basis functions \cite{Mehl.2024,Ataka.2016,Loo.2019,Lobaton.2013,Zhang.2026,Abdelaziz.2023}. 
    To account for unmodeled effects in the latter, limited model fidelity is often compensated for by estimating unknown parameters as slack variables. 
    For this, a typical approach is to estimate the parameters alongside the states in a nonlinear Kalman filter \cite{Mehl.2024}, using a random walk assumption. 
    Notably, while all these works provide state estimates that might be used for control, they rely on a fixed, often nominal, system model. 
    This hampers accurate multi-step predictions---a prerequisite for predictive control. 
    Instead, an interesting approach, which we follow in this paper, is to perform online model learning alongside state estimation. 

	

	
	

	


	\section{Problem Formulation}\label{sec:problem}

    Consider the situation when a nominal system model---here, a dynamic constant-curvature soft robot model---and sequentially arriving noisy output measurements are available. 
    Given this, we aim to simultaneously estimate the unknown latent state representing the robot's pose and velocity, and learn an augmented system model to improve the overall prediction accuracy. 
	Formally, consider a nonlinear stochastic discrete-time state-space system
	\begin{equation}\label{eq:problem}
		\begin{aligned}
			 \boldsymbol{x}_{t+1} &=  \boldsymbol{f}_{\mathrm{nom}}( \boldsymbol{x}_{t},  \boldsymbol{u}_{t}, k_{\mathrm{bend}}( \boldsymbol{x}_{t})) +  \boldsymbol{\omega}_t \, ,\\
			 \boldsymbol{y}_{t} &=  \boldsymbol{h}_{\mathrm{nom}}( \boldsymbol{x}_{t},  \boldsymbol{u}_{t}, k_{\mathrm{bend}}( \boldsymbol{x}_{t})) +  \boldsymbol{e}_t \, ,
		\end{aligned}
	\end{equation}
	where the state transition function $ \boldsymbol{f}_{\mathrm{nom}}: \mathbb{R}^{n_{x}} \times \mathbb{R}^{n_{u}} \times \mathbb{R} \rightarrow \mathbb{R}^{n_{x}}$, and the measurement function $ \boldsymbol{h}_{\mathrm{nom}}: \mathbb{R}^{n_{x}} \times \mathbb{R}^{n_{u}} \times \mathbb{R} \rightarrow \mathbb{R}^{n_{y}}$ together constitute the nominal model, which is based on a constant-curvature assumption. 
    The predictive accuracy of the nominal model is limited due to various modeling assumptions. 
    The formulation \eqref{eq:problem} accounts for this by augmenting the nominal system equations with a state-dependent bending stiffness model ${k}_{\mathrm{bend}} : \mathbb{R}^{n_{x}} \rightarrow \mathbb{R}$, which is to be learned. 

    In both the state transition and measurement equation, we assume additive Gaussian noise $ \boldsymbol{\omega}_t\,\sim\,\mathcal{N}(\boldsymbol{\omega}_t | \boldsymbol{0}, \boldsymbol{\Sigma}_{\omega})$ and $\boldsymbol{e}_t\,\sim\, \mathcal{N}(\boldsymbol{e}_t | \boldsymbol{0}, \boldsymbol{\Sigma}_e)$ with covariances $\boldsymbol{\Sigma}_{\omega}$ and $\boldsymbol{\Sigma}_e$, respectively. 
    The inputs that drive the robot are pneumatic pressures, denoted by $ \boldsymbol{u}_t \in \mathbb{R}^{n_{u}}$ at time step $t$. 
    The outputs $ \boldsymbol{y}_t \in \mathbb{R}^{n_{y}}$ are base reaction forces and torques, measured using a force-torque sensor. 
    The sensor setup is depicted in Figure~\ref{fig:problem_setting}.

    Given the input-output data and the nominal model, our goal is to estimate the unknown hidden robot states $ \boldsymbol{x}_t \in \mathbb{R}^{n_{x}}$, \ie the the robot's pose and velocity, while simultaneously learning the nonlinear bending stiffness model ${k}_{\mathrm{bend}} ( \boldsymbol{x}_t)$. 

    \section{Key Idea}\label{sec:key_idea}
    
    Learning a bending stiffness model \textit{in the context} of the constant-curvature approximation $ \boldsymbol{f}_{\mathrm{nom}}$ \& $ \boldsymbol{h}_{\mathrm{nom}}$ is challenging for two reasons. 
    First, while the nominal model itself might be known, it cannot be assumed to be invertible. 
    This prevents us from learning the bending stiffness with ``pseudo-measurements'', constructed using a coordinate transformation $\left.  \boldsymbol{f}_{\mathrm{nom}}^{-1}( \boldsymbol{x}_{t+1}) \right|_{ \boldsymbol{x}_t,  \boldsymbol{u}_t} \approx {k}_{\mathrm{bend}} ( \boldsymbol{x}_t)$. 
    Second, the soft robot's states, which are regression inputs to ${k}_{\mathrm{bend}} ( \boldsymbol{x}_t)$, are unknown and need to be estimated. 

    We tackle these problems by adopting a sampling-based estimation strategy. 
    This avoids inverting the nominal model while enabling state inference. 
    In particular, we estimate unknown states using a particle filter---marginalized over the bending stiffness model parameters---that samples \textit{forward} through the entire model \eqref{eq:problem}. 
    Sequentially, the bending stiffness \ac{gp} model is learned using the posterior state and bending stiffness estimates (see Figure\,\ref{fig:problem_setting}).

    The gray-box system identification tool has been proposed in our recent work \cite{Volkmann.2025}, and we exploit it here to simultaneously perform state estimation and online model learning in the nonlinear soft robotic system. 
    Therefore, the hierarchical soft robot model is introduced in Section\,\ref{sec:model}, and the estimation scheme is revisited in Section\,\ref{sec:methods}. 
    We thoroughly test the algorithm and the learned model using real-world data in Section\,\ref{sec:results} and draw conclusions in Section\,\ref{sec:conclusion}.
    
	\section{Soft Robot Model}\label{sec:model}
    In this section, we present a suitable model structure in accordance with \eqref{eq:problem}. 
    First, the nominal system equations based on constant-curvature assumptions are briefly revisited in Section\,\ref{sec:model_cc}. 
    Section\,\ref{sec:model_stiffness} is dedicated to constructing a \ac{gp}-based bending stiffness model for online learning. 

	\subsection{Nominal Constant-Curvature Model}\label{sec:model_cc}

    For the nominal system equations, we adopt the soft robot model presented in \cite{Mehl.2022,Mehl.2024}. 
    At its core, the model relies on a constant-curvature kinematics assumption to approximate the arc of a soft robot. 
    Therefore, the robot state is fully described by its elongation $\delta L$ relative to a reference length, as well as its relative coordinates $\delta x$ and $\delta y$ from the robot's base to the robot's tip \cite{DellaSantina.2020}. 
    Thus, we have the generalized coordinates $ \boldsymbol{q}^{\top} = [\delta x, \delta y, \delta L]$, $ \boldsymbol{q} \in \Omega \subset \mathbb{R}^{n_q}$, and $n_q = 3$. 
    The full state vector comprises the generalized coordinates and their time derivatives, \ie $ \boldsymbol{x}^{\top} = [ \boldsymbol{q}^{\top}, \dot{ \boldsymbol{q}}^{\top}] \in \mathbb{R}^{n_x}$, and $n_x=6$. 
    The base reaction forces and torques $ \boldsymbol{y} \in \mathbb{R}^{6}$, relevant for modeling the measurement function $ \boldsymbol{h}_{\mathrm{nom}}$, are incorporated using a floating base approach. 
    Finally, using Lagrangian equations of the second kind, the state dynamics $ \boldsymbol{f}_{\mathrm{nom}}$ are constructed. 
    For a detailed derivation of the model, we refer to \cite{Mehl.2022,Mehl.2024}.

    The model relies on three main assumptions: a constant robot curvature, a concentrated robot mass, and a stiffness model that is affine in the generalized coordinates. 
    The latter approximates the restoring force $\boldsymbol{\tau}_{\mathrm{bend}}$ to bending the soft robot in one direction, \ie we have $\tau_{\mathrm{bend},x}\,=\,{k}_{\mathrm{bend}} \delta x $ for the coordinate $\delta x$. 
    The modeling assumptions significantly limit the model's predictive accuracy. 
    In fact, due to model simplifications, the state estimation scheme proposed in \cite{Mehl.2024} necessitates estimating the parameter ${k}_{\mathrm{bend}}$ as an auxiliary slack variable to enable satisfactory filter performance. 
    Instead, we augment the nominal system equations with a state-dependent bending stiffness model ${k}_{\mathrm{bend}} ( \boldsymbol{x})$ to account for lacking model fidelity and to enable accurate multi-step predictions. 

	\subsection{Bending Stiffness Model}\label{sec:model_stiffness}

    We model the bending stiffness as a state-dependent function to be learned. 
    The function is assumed to only depend on the current pose of the robot, \ie ${k}_{\mathrm{bend}}( \boldsymbol{x}) \equiv {k}_{\mathrm{bend}}( \boldsymbol{q})$. 
    
	To obtain a data-efficient model for online learning, we approximate the bending stiffness using a zero-mean \ac{gp}, \ie 
    \begin{equation}\label{eq:gp_bending_stiffness}
        {k}_{\mathrm{bend}}( \boldsymbol{x}) \sim  \mathcal{GP}\left( 0, \kappa(\boldsymbol{q}, \boldsymbol{q}') \right) \, ,
    \end{equation}
    with covariance (kernel) function $\kappa(\boldsymbol{q}, \boldsymbol{q}')$ \cite{Rasmussen.2005}. 
	A common drawback of \acp{gp} is that their computational complexity and memory requirements scale poorly with the number of training data points. 
	However, for efficient online learning, it is desirable that the bending stiffness model features \textit{(i)} a computational complexity that is independent of the training data dimension and \textit{(ii)} a linear representation to facilitate closed-form marginalization (see Section~\ref{sec:methods}). 
	
	Importantly, the reduced-rank \ac{gp} presented in \cite{Solin.2020} exhibits these properties, making it a popular choice when it comes to online inference and learning \cite{Berntorp.2021,Volkmann.2025}. 
    Please note that we revisit the \ac{gp} construction only briefly, based on our prior work \cite{Volkmann.2025,Ewering.2025}, while applying it to bending stiffness modeling. 
    
	Loosely speaking, the reduced-rank \ac{gp} \cite{Solin.2020} approximates the kernel by a finite-dimensional eigenfunction expansion that encodes its spectral density $S$ in the frequency domain. 
	The chosen covariance function is described by
	\begin{equation}\label{eq:kernel_approx}
		\kappa(\boldsymbol{q}, \boldsymbol{q}') \approx \sum_{m=1}^{M} S(\sqrt{\varrho_m}) \phi_m(\boldsymbol{q}) \phi_m(\boldsymbol{q}'),
	\end{equation}
	where $\phi_m : \Omega \rightarrow \mathbb{R}$ are eigenfunctions of the Laplace operator on the domain $\Omega = [-L_1, L_1] \times \dots \times [-L_{n_q}, L_{n_q}]$, and $\varrho_m$ are the corresponding eigenvalues. 
	For this rectangular domain, the eigenfunctions have a closed form, that is
	\begin{align}\label{eq:basis_functions}
		\phi_{m}(\boldsymbol{q}) &= \prod_{i=1}^{n_q} {{L_i^{-\frac{1}{2}}}} \sin \left(\frac{\pi j_{m,i}\left(q_{i}+L_i\right)}{2 L_i}\right) , \:
        \varrho_{m} = \sum_{i=1}^{n_q} \frac{\pi^2 j_{m,i}^2}{4 L_i^2} , \nonumber
	\end{align}
	where the indices $j_{m,i}$ determine the frequency of the corresponding eigenfunction \cite{RiutortMayol.2023}. 
	Predictions of ${k}_{\mathrm{bend}}(\boldsymbol{x}) \sim \mathcal{GP}(0, \kappa(\boldsymbol{q}, \boldsymbol{q}'))$ can be performed with the reduced-rank \ac{gp} at computational complexity $\mathcal{O}(M)$ using the following simple basis function expansion
	\begin{equation}\label{eq:rr_gp_pred}
		\hat{k}_{\mathrm{bend}}( \boldsymbol{x}) = \sum_{m=1}^{M} a_m \phi_m(\boldsymbol{q}) =  \boldsymbol{a}^\top   \boldsymbol{\phi}(\boldsymbol{q}) \, ,
	\end{equation}
	where the vector-valued function 
    \begin{equation}
         \boldsymbol{\phi}(\boldsymbol{q}) = [\phi_1(\boldsymbol{q}), \hdots, \phi_M(\boldsymbol{q})]^\top \, ,
    \end{equation}
    and the weights $ \boldsymbol{a} = [a_1, \hdots, a_M]^\top$ follow a density $\mathcal{N} (  \boldsymbol{0},  \boldsymbol{V})$, with $ \boldsymbol{V} = \mathrm{diag}(S(\sqrt{\varrho_1}),\hdots,S(\sqrt{\varrho_M}))$. 
    
	To estimate the current bending stiffness (and using it for parameter learning) at a given sample time $t$, we model observations ${k}_{t} \in \mathbb{R}$ of $\hat{k}_{\mathrm{bend}}( \boldsymbol{x})$ as
	\begin{equation}\label{eq:probab_model2}
		\begin{aligned}
			{k}_t  =  \left.  \hat{k}_{\mathrm{bend}} (\boldsymbol{x}) \right|_{\boldsymbol{x} = \boldsymbol{x}_t} + {v}_t \, , \qquad {v}_t \sim \mathcal{N}({v}_t \mid {0}, \sigma^2 ) \, .
		\end{aligned}
	\end{equation}
	For conjugacy reasons and to describe the model \eqref{eq:rr_gp_pred} and \eqref{eq:probab_model2} with a single density \cite{Svensson.2017,Berntorp.2021}, we express the \ac{gp} prior as a zero-mean multivariate normal $\mathcal{N}( \boldsymbol{a}  \mid   \boldsymbol{0},\sigma^2  \boldsymbol{V})$, where the scale $\sigma^2 $ reflects the noise, and the covariance $ \boldsymbol{V}$ encodes the spectral density of the chosen kernel. 
	While any isotropic kernel can be employed, we resort to approximating a common squared exponential kernel
	\begin{align}
		\kappa_{\mathrm{se}}(\boldsymbol{q}, \boldsymbol{q}') &= \sigma_f^2 \exp\left( -\frac{\lVert \boldsymbol{q} - \boldsymbol{q}' \rVert^2}{2\ell^2} \right) \, , 
    \end{align}
    with Euclidean norm $\norm{\cdot}$ and the spectral density
    \begin{align}\label{eq:spectral_density}
        S_{\mathrm{se}}(\omega) &= \sigma_f^2 (2 \pi \ell^2)^{n_q/2} \exp \left(-\frac{ \ell^2 \omega^2}{2}\right) \, .
	\end{align}
	The kernel hyperparameters $ \boldsymbol{\vartheta} = \{\sigma_f^2, \ell\}$ are scale $\sigma_f^2$ and length scale $\ell$. 
	
	The noise parameter $\sigma^2$ in \eqref{eq:probab_model2} is unknown and must be estimated along with the basis function coefficients $ \boldsymbol{a}$. 
	A typical approach for online noise estimation is to set an inverse Gamma prior $\mathcal{IG}(\sigma^2 \mid  {\psi}, \nu)$ on $\sigma^2$ \cite{Ozkan.2013, Murphy.2007},
	with scale $\psi$ and degrees of freedom $\nu$. 
    The overall reduced-rank \ac{gp} prior for the parameters $ \boldsymbol{\theta} = \{ \boldsymbol{a}, \sigma^2\}$ is then a normal inverse Gamma density ($\mathcal{NIG}$), \ie
	\begin{equation} \label{eq:MdlStr_MNIW_Prior}
		\begin{aligned}
			 \boldsymbol{a}, \sigma^2 \sim & \, \mathcal{NIG}( \boldsymbol{a}, \sigma^2 \mid   \boldsymbol{0},  \boldsymbol{V}, {\psi}, \nu)  \\
            & \qquad  =\mathcal{N}( \boldsymbol{a} \mid  \boldsymbol{0}, \sigma^2  \boldsymbol{V}  )\mathcal{IG}(\sigma^2 \mid  {\psi}, \nu) \, .
		\end{aligned}
	\end{equation}
    Please note that using the prior \eqref{eq:MdlStr_MNIW_Prior} enables efficient marginalization and recursive closed-form computation of the parameter posterior, which is convenient for online learning \cite{Berntorp.2021}.

	For learning the bending-stiffness model \textit{online}, we consider time-varying parameters $ \boldsymbol{\theta}_t$ at time step $t$, which must be estimated from sequentially arriving base reaction measurements $ \boldsymbol{y}_t$. 
    Moreover, the \ac{gp} kernel hyperparameters $ \boldsymbol{\vartheta}_t$ should be learned as well. 
    
	\section{Online Inference and Learning}\label{sec:methods}

    In this section, we revisit the gray-box system identification tool---recently introduced in \cite{Volkmann.2025}---for simultaneously estimating hidden states and learning model parameters. 
    In particular, we approximate the joint posterior density\footnote{To retain a concise notation, dependence on the system inputs $ \boldsymbol{u}_t$ is not explicitly written in this section.}
	\begin{align} \label{eq:MdlStr_posterior}
		p( \boldsymbol{x}_{0:t}, k_{0:t},  \boldsymbol{\theta}_{t} \mid  \boldsymbol{y}_{0:t}) 
		&= \underbrace{p( \boldsymbol{\theta}_{t} \mid  \boldsymbol{x}_{0:t}, k_{0:t})}_{\text{posterior (i)}} \underbrace{p( \boldsymbol{x}_{0:t}, k_{0:t} \mid  \boldsymbol{y}_{0:t})}_{\text{posterior (ii)}} \, , \nonumber
	\end{align}
	and construct a marginalized \ac{pf} by recursively sampling from two simpler densities in each time step $t$ \cite{Ozkan.2013,Berntorp.2021}. 
    Specifically, we compute the posterior (i) in closed form by fusing the likelihood of the current hidden states with the conjugate \ac{gp} parameter prior in \eqref{eq:MdlStr_MNIW_Prior} (see Section\,\ref{sec:methods_parameters}). 
    Vice versa, we construct a particle approximation of posterior (ii) using a \ac{pf}, marginalized over the current model parameter density (see Section\,\ref{sec:methods_states}). 
    
    Importantly, both closed-form posterior and parameter marginalization rely on the linear-in-its-parameters bending stiffness model, defined in Section\,\ref{sec:model_stiffness}, and we refer to the \ac{pf} construction as a ``marginalized \ac{pf}'' in the following.  
    
    Beyond our prior work \cite{Volkmann.2025}, we take inspiration from \cite{Berntorp.2021} and adopt a learning scheme for the \ac{gp} kernel hyperparameters $ \boldsymbol{\vartheta}$ (see Section\,\ref{sec:methods_hyperparameters}). 
    The overall procedure for simultaneous state estimation and online model learning is given in Algorithm\,\ref{alg:online_RBPF}, and the key idea is summarized in Figure\,\ref{fig:problem_setting}.

    \subsection{Online Model Learning}\label{sec:methods_parameters}
    Assume, for now, that state estimates $ \boldsymbol{x}_{t}$ and bending stiffness estimates $k_{t}$ are provided sequentially by the marginalized \ac{pf}. 
    Given these, let us now derive recursive equations to compute the parameter posterior $p( \boldsymbol{\theta}_{t} \mid  \boldsymbol{x}_{0:t}, k_{0:t})$. 
    We employ a scheme for online noise parameter learning with marginalized \acp{pf}, inspired by \cite{Ozkan.2013} and adopted for model learning in \cite{Berntorp.2021,Volkmann.2025}. 
    To retain conciseness, we state only the main steps and refer to \cite{Ozkan.2013} for further details. 
    To start with, the current parameter posterior can be decomposed as
		\begin{equation}\label{eq:online_param_posterior}
			\begin{aligned}
				&p( \boldsymbol{\theta}_{t} \mid  \boldsymbol{x}_{0:t}, k_{0:t}) \propto  \\
				&\quad p( \boldsymbol{x}_{t}, k_{t}  \mid   \boldsymbol{\theta}_{t},  \boldsymbol{x}_{t-1}, k_{t-1}) p( \boldsymbol{\theta}_{t} \mid  \boldsymbol{x}_{0:t-1}, k_{0:t-1}) \, ,
			\end{aligned}
		\end{equation}
		using Bayes' rule. 
        Since the constant-curvature model \eqref{eq:problem} is independent of $ \boldsymbol{\theta}_t$, the likelihood is given by
		\begin{align}\label{eq:online_parameter_decomposition}
				& p( \boldsymbol{x}_{t}, k_{t} \mid   \boldsymbol{\theta}_{t},  \boldsymbol{x}_{t-1}, k_{t-1}) \nonumber \\
                & \qquad \qquad = p( k_{t} \mid  \boldsymbol{x}_{t},  \boldsymbol{\theta}_{t} ) p( \boldsymbol{x}_{t} \mid   \boldsymbol{x}_{t-1}, k_{t-1}) \, .
		\end{align}
        In \eqref{eq:online_parameter_decomposition}, we see that the likelihood follows the normal density
        \begin{equation}
            p( k_{t} \mid  \boldsymbol{x}_{t},  \boldsymbol{\theta}_{t} ) = \mathcal{N}( {k}_t \mid  \boldsymbol{a}_t^\top   \boldsymbol{\phi}(\boldsymbol{q}_t), \sigma_t^2 ) \, ,
        \end{equation}
        up to a scaling, resulting from $p( \boldsymbol{x}_{t} |  \boldsymbol{x}_{t-1}, k_{t-1})$. 
		Given the conjugate normal-inverse-Gamma prior \eqref{eq:MdlStr_MNIW_Prior} and the likelihood \eqref{eq:online_parameter_decomposition}, the parameter posterior $p( \boldsymbol{\theta}_{t} |  \boldsymbol{x}_{0:t}, k_{0:t})$ is again a normal-inverse-Gamma density \cite{Ozkan.2013,Berntorp.2021}
        \begin{align}\label{eq:parameter_posterior}
            p( \boldsymbol{\theta}_t \mid  \boldsymbol{x}_{0:t}, k_{0:t}) = \mathcal{NIG}( \boldsymbol{a}, {\sigma}^2 \mid   \boldsymbol{m}_{t \mid t},  \boldsymbol{V}_{t \mid t}, {\psi}_{t \mid t}, \nu_{t \mid t}) \, ,
        \end{align}
        where $\boldsymbol{m}_{t \mid t}$, $\boldsymbol{V}_{t \mid t}$, ${\psi}_{t \mid t}$, and $\nu_{t \mid t}$ are the posterior density parameters, given the estimates $\boldsymbol{x}_t$ and $k_t$. 
        Conveniently, the posterior \eqref{eq:parameter_posterior} can be recursively computed in closed form by summation of the priors' sufficient statistics and new statistics obtained from the current estimates $ \boldsymbol{x}_{t}$, $k_{t}$ \cite{Berntorp.2021,Volkmann.2025}. 
        Specifically, the posterior density is characterized by
    	\begin{equation} \label{eq:MdlStr_mniw_suffstat2para}
    		\begin{aligned}
    			 \boldsymbol{m}_{t \mid t} &=  \boldsymbol{r}_{1,t \mid t}^{-1}  \boldsymbol{s}_{1,t \mid t} \, , \qquad &  \boldsymbol{V}_{t \mid t} &=  \boldsymbol{r}_{1,t \mid t}^{-1} \, ,\\
    			{\psi}_{t \mid t} &= {s}_{2,t \mid t} -  \boldsymbol{s}_{1,t \mid t}^{\top} \boldsymbol{r}_{1,t \mid t}^{-1}  \boldsymbol{s}_{1,t \mid t} \, , \: & \nu_{t \mid t} &= r_{2,t \mid t} \, ,
    		\end{aligned}
    	\end{equation}
        with updated statistics $ \boldsymbol{\eta}_{t \mid t} = \{ \boldsymbol{s}_{1,t \mid t}, {s}_{2,t \mid t},  \boldsymbol{r}_{1,t \mid t}, {r}_{2,t \mid t} \}$, \ie
        \begin{equation}\label{eq:online_meas_update}
            \begin{aligned}
                 \boldsymbol{s}_{1,t \mid t} &=  \boldsymbol{s}_{1,t \mid t-1} +   \boldsymbol{\phi}( \boldsymbol{q}_{t})k_{t} \, , \\
                {s}_{2,t \mid t} &= {s}_{2,t \mid t-1} +  k_{t}^2 \, , \\
                 \boldsymbol{r}_{1,t \mid t} &=  \boldsymbol{r}_{1,t \mid t-1} +   \boldsymbol{\phi}( \boldsymbol{q}_{t})  \boldsymbol{\phi}( \boldsymbol{q}_{t})^{\top} \, , \\
                r_{2,t \mid t}      &= r_{2,t \mid t-1} + 1 \, .
            \end{aligned}
        \end{equation}
        To incorporate the \ac{gp} prior \eqref{eq:MdlStr_MNIW_Prior}, the initial $\mathcal{NIG}$ distribution parameters are set as
    	\begin{equation} \label{eq:MdlStr_mniw_suffstat2para_initial}
    		\begin{aligned}
    			 \boldsymbol{m}_{0} &=  \boldsymbol{0} \, , \quad   \boldsymbol{V}_{0} =  \boldsymbol{V} \, , \quad {\psi}_{0} &= {\psi} \, , \quad \nu_{0} = \nu \, ,
    		\end{aligned}
    	\end{equation}
        which corresponds to the initial sufficient statistics $\boldsymbol{\eta}_{0}$ by the inverse relationship of \eqref{eq:MdlStr_mniw_suffstat2para}.
        
		If the parameters $ \boldsymbol{\theta}_t$ are known to vary over time, exponential forgetting can be employed to discard old information contained in the sufficient statistics. 
        We introduce forgetting in the time update step of the statistics $ \boldsymbol{\eta}_{t \mid t-1}$, that is,
		\begin{equation}\label{eq:online_time_update}
			\begin{aligned}
				 \boldsymbol{s}_{i,t \mid t-1} &= \gamma  \boldsymbol{s}_{i,t-1 \mid t-1} \, , \quad \boldsymbol{r}_{i,t \mid t-1} = \gamma  \boldsymbol{r}_{i,t-1 \mid t-1} \, , \quad i=1,2\,, 
			\end{aligned}
		\end{equation}
        where $0 \leq \gamma \leq 1$ is the so-called forgetting factor \cite{Ljung.1999}. 

	\subsection{State Estimation}\label{sec:methods_states}
        Given the current parameter density, defined by the statistics $ \boldsymbol{\eta}_{t \mid t-1}$, we approximate the filtering density
		\begin{equation}\label{eq:online_state_posterior}
			\begin{aligned}
				p( \boldsymbol{x}_{0:t}, k_{0:t} \mid  \boldsymbol{y}_{0:t}) \approx \sum_{i=1}^{N} w_{t}^i \delta_{\{  \boldsymbol{x}_{0:t}^i, k_{0:t}^i\}}( \boldsymbol{x}_{0:t}, k_{0:t}),
			\end{aligned}
		\end{equation}
		as a set of $N$ weighted particles, each representing a trajectory of state and bending stiffness values. 
        In \eqref{eq:online_state_posterior}, the Dirac delta mass centered at $ \boldsymbol{z}$ is denoted by $\delta_{ \boldsymbol{z}} (\cdot)$. 
        The importance weight of each particle $i$ at time step $t$ is denoted $w_t^{i}$, and it is associated with trajectories $ \boldsymbol{x}_{0:t}^i, k_{0:t}^i$.

        The trajectories are obtained by recursively applying the three steps that define a standard \ac{smc} algorithm: sampling, weighting, and resampling \cite{Sarkka.2023}. 
        Specifically, we use an auxiliary \ac{pf} \cite{Sarkka.2023} to increase the sample efficiency.  
        In the sampling step, state and bending stiffness samples are drawn from a tractable proposal distribution $\pi ( \boldsymbol{x}_t, k_t  \mid   \boldsymbol{x}_{0:t-1}, k_{0:t-1}, \boldsymbol{y}_{0:t})$. 
        Then, the weights $\{w_{t}^i\}_{i=1}^{N}$ are updated according to 
		\begin{equation}\label{eq:online_weight_update}
			w^{i}_{t} \propto \frac{p\left( \boldsymbol{y}_{t} \mid   \boldsymbol{x}^{i}_{0:t}, k^{i}_{0:t}\right)}{p\left( \boldsymbol{y}_{t} \mid  \tilde{ \boldsymbol{x}}^{a_t^{i}}_{t},\tilde{k}^{a_t^{i}}_{t} \right)} = \frac{\mathcal{N}\left( \boldsymbol{y}_t  \mid   \boldsymbol{h}\left({ \boldsymbol{x}}_t^{i}, { \boldsymbol{u}}_t, {k}_t^{i}\right) , \boldsymbol{\Sigma}_e\right)}{\mathcal{N}\left(  \boldsymbol{y}_t  \mid   \boldsymbol{h}\left( \tilde{ \boldsymbol{x}}_t^{a_t^{i}}, { \boldsymbol{u}}_t, \tilde{k}_t^{a_t^{i}}\right) , \boldsymbol{\Sigma}_e \right)} \, .
		\end{equation}
        
        While the weighting step is easily computed using the drawn state and bending stiffness values, the sampling procedure requires more attention to account for our current model parameter estimate. 
        In fact, generating samples for the state and bending stiffness at time step $t$, given the previous trajectory estimates $ \boldsymbol{x}_{0:t-1}$ and $k_{0:t-1}$, requires drawing from the state transition density $p( \boldsymbol{x}_{t},k_{t} |  \boldsymbol{x}_{0:t-1}, k_{0:t-1})$. 
        This density is composed as
		\begin{equation}\label{eq:online_marginalization}
			\begin{aligned}
				&p( \boldsymbol{x}_{t},k_{t} \mid  \boldsymbol{x}_{0:t-1}, k_{0:t-1}) \\
				&= p( \boldsymbol{x}_{t} \mid  \boldsymbol{x}_{t-1}, k_{t-1}) p(k_{t} \mid  \boldsymbol{x}_{t}, \boldsymbol{\eta}_{t \mid t-1})\\
				&= p( \boldsymbol{x}_{t} \mid  \boldsymbol{x}_{t-1}, k_{t-1}) \int p(k_{t} \mid  \boldsymbol{x}_{t},  \boldsymbol{\theta}) p( \boldsymbol{\theta} \mid   \boldsymbol{\eta}_{t \mid t-1}) \mathrm{d} \boldsymbol{\theta} \, ,
			\end{aligned}
		\end{equation}
        where the first term on the right-hand side is defined by \eqref{eq:problem}. 
        The second term in \eqref{eq:online_marginalization}, \ie the predictive bending stiffness density $p(k_{t} \mid  \boldsymbol{x}_{t}, \boldsymbol{\eta}_{t \mid t-1})$, is obtained by integrating out the parameters $ \boldsymbol{\theta}$ from the hierarchical models 
		\begin{equation}\label{eq:online_hierarchical_sampling}
			\begin{aligned}
				k_{t} &\sim \mathcal{N}\left(k_{t} \mid  \boldsymbol{a}^{\top}  \boldsymbol{\phi}( \boldsymbol{x}_{t}), {\sigma}^2 \right),\\
				 \boldsymbol{a}, {\sigma}^2 & \sim \mathcal{NIG}( \boldsymbol{a},  \boldsymbol{\sigma}^2 \mid { \boldsymbol{m}}_{t \mid t-1}^{i}, { \boldsymbol{V}}_{t \mid t-1}^{i}, {{\psi}}_{t \mid t-1}^{i}, {\nu}_{t \mid t-1}^{i}) \, ,
			\end{aligned}
		\end{equation}
        for each particle $i$. 
        Therefore, each particle carries its own bending stiffness model, represented by an individual set of statistics $ \boldsymbol{\eta}_{t \mid t-1}^{i}$. 
        Conveniently, due to the conjugate prior configuration in \eqref{eq:online_hierarchical_sampling}, each predictive density $p(k_{t} |  \boldsymbol{x}_{t}, \boldsymbol{\eta}_{t \mid t-1})$ follows a Student-t distribution ($\mathcal{T}$) \cite{Volkmann.2025} with
        \begin{equation}\label{eq:online_predictive_distribution}
                p(k_{t} \mid  \boldsymbol{x}_{t}, \boldsymbol{\eta}_{t \mid t-1}) = \mathcal{T}(k_{t} \mid   {\rho}, {{\mu}}, {{\Lambda}} ) \,,
        \end{equation}
        and distribution parameters
        \begin{equation}\label{eq:online_predictive_distribution_parameters}
            \begin{aligned}
                 {\rho} &=  {\nu}_{t \mid t-1} \, , &  {{\mu}} &=  { \boldsymbol{m}_{t \mid t-1}}^{\top} \boldsymbol{\phi}( \boldsymbol{x}_t) \, ,\\
                 {{\Lambda}} &= \frac{ {\xi} + 1}{ {\xi}  {\rho}}  {{\psi_{t \mid t-1}}}\, ,  &  {\xi} &= 1/\left( \boldsymbol{\phi}( \boldsymbol{x}_t)^{\top}  { \boldsymbol{V}_{t \mid t-1}} \boldsymbol{\phi}( \boldsymbol{x}_t)\right) \, .
            \end{aligned}
        \end{equation}
        This enables efficient sampling and, hence, state estimation while---through parameter marginalization---respecting the different bending stiffness models. 
        For further details on the algorithm, we refer to \cite{Volkmann.2025}.

	\begin{algorithm}[tb]
    \begin{minipage}{\linewidth} 
		\caption{Marginalized \ac{pf} for online state estimation and model learning (for all $i=1,\dots,N$)}
		\label{alg:online_RBPF}
		\textbf{Initialize:} Data $ \boldsymbol{y}_{0:T}$, prior $p( \boldsymbol{\theta})$, hyperparameters $\boldsymbol{\vartheta}^{i}$, particles $  \boldsymbol{x}_{0}^{i},k_{0}^{i}  \sim p(  \boldsymbol{x}_0,k_0  | \boldsymbol{\theta})$, weights $w^i_{0} = 1/N$.
		\begin{algorithmic}[1]
			\For{$t = 1,\dots,T$}
			\State Statistics time update $ \boldsymbol{\eta}^i_{t \mid t-1} \leftarrow  \boldsymbol{\eta}^i_{t-1 \mid t-1}$. \Comment{by \eqref{eq:online_time_update}}
			\State Comp. auxiliary states $\tilde{ \boldsymbol{x}}_t^{i} =  \boldsymbol{f}_{\mathrm{nom}} ( \boldsymbol{x}_{t-1}^{i}, \boldsymbol{u}_{t-1}, k_{t-1}^{i})$.
			\State Compute first-stage weights 
			\Statex \hspace{4mm} $\tilde{\lambda}^{i} = w_{t-1}^{i} \mathcal{N}( \boldsymbol{y}_t  \mid   \boldsymbol{h}_{\mathrm{nom}}(\tilde{ \boldsymbol{x}}_t^{i}, \boldsymbol{u}_{t}, k_{t}^{i}) , \boldsymbol{\Sigma}_e)$ 
			\Statex \hspace{4mm} and normalize $\lambda^{i} = \tilde{\lambda}^{i} / \sum_{j=1}^N \tilde{\lambda}^{j} $.
			\State Resample\footnote{We describe the resampling step using a categorical distribution $\mathcal{C}$ to express sampling particle indices from a discrete probability mass function, spanned by the weights $\{\lambda^{i}\}_{i=1}^{N}$.} $a_t^{i} \sim \mathcal{C}(\{\lambda^{i}\}_{i=1}^{N})$.
            \If{\ac{gp} hyperparameter learning}
            \State Sample hyperparameters $\boldsymbol{\vartheta}_{t}^{i}$. \Comment{by \eqref{eq:hyperparameter_sampling}}
            \State Update prior $\boldsymbol{V}^{i}$. \Comment{by \eqref{eq:spectral_density}}
            \EndIf
			\State Draw $ \boldsymbol{x}_t^{i} \sim \mathcal{N}( \boldsymbol{x}_t \mid  \boldsymbol{f}_{\mathrm{nom}} ( \boldsymbol{x}_{t-1}^{a_t^{i}},\boldsymbol{u}_{t-1}, k_{t-1}^{a_t^{i}}),  \boldsymbol{\Sigma}_\omega)$.
			\State Draw $k_t^{i} \sim p(k_t  \mid   \boldsymbol{x}_t^{i},  \boldsymbol{\eta}^{a_t^{i}}_{t \mid t-1})$. \Comment{by \eqref{eq:online_predictive_distribution}}
			\State Statistics measurement update and
			\Statex \hspace{4mm} resampling $ \boldsymbol{\eta}^i_{t \mid t} \leftarrow  \boldsymbol{\eta}^{a_t^{i}}_{t \mid t-1}$. \Comment{by \eqref{eq:online_meas_update}}
			\State Compute weights ${{w}}_t^{i}$ and normalize. \Comment{by \eqref{eq:online_weight_update}}
			\EndFor
		\end{algorithmic}
        \end{minipage} 
	\end{algorithm}
    
	\subsection{Hyperparameter Learning}\label{sec:methods_hyperparameters}

    While it is often intractable to learn all hyperparameters online \cite{Berntorp.2021}, the \ac{gp} kernel hyperparameters \textit{can} actually be determined in our state estimation and model learning setup. 
    This has the potential to further align the learned \ac{gp} model with the training data and enable more accurate predictions, while requiring less expert knowledge. 
    
    For learning the kernel hyperparameters $\boldsymbol{\vartheta}$, we follow the lines of \cite{Berntorp.2021} and model the hyperparameter evolution with a random walk, \ie
    \begin{equation}\label{eq:hyperparameter_sampling}
        \boldsymbol{\vartheta}_{t+1} = \boldsymbol{\vartheta}_{t} + \boldsymbol{\zeta}_t \, , \qquad \boldsymbol{\zeta}_t \sim \mathcal{N} \left( \boldsymbol{0}, \boldsymbol{\Sigma}_{\zeta} \right) \, ,
    \end{equation}
    where $\boldsymbol{\Sigma}_{\zeta}$ is a diagonal matrix. 
    Without the hyperparameter learning scheme, the \ac{gp} prior is fixed and shared across all particles. 
    In contrast, to learn the hyperparameters, each particle $i$ carries its own hyperparameter set $\boldsymbol{\vartheta}_{t}^{i}$ and corresponding \ac{gp} prior, represented by the covariance matrix $\boldsymbol{V}^{i}$. 
    
    During each time step $t$, the particle hyperparameters $\boldsymbol{\vartheta}_{t}^{i}$ are updated using \eqref{eq:hyperparameter_sampling}. 
    Then, a new matrix $\boldsymbol{V}^{i}$ for the $\mathcal{NIG}$ prior \eqref{eq:MdlStr_MNIW_Prior} is generated. 
    Through \eqref{eq:MdlStr_mniw_suffstat2para_initial}, this affects two steps in the marginalized \ac{pf}: \textit{(i)} the parameter posterior \eqref{eq:parameter_posterior}, and \textit{(ii)} the predictive sampling of the bending stiffness \eqref{eq:online_predictive_distribution}. 
    Thus, along the time dimension, suitable hyperparameters are learned by successive sampling, weighting, and resampling $\boldsymbol{\vartheta}_t$.

    \subsection{Computational Complexity}\label{sec:results_computational_aspects}
    The majority of operations in Algorithm\,\ref{alg:online_RBPF} are performed for each particle. 
    Due to the auxiliary \ac{pf}, the maps $\boldsymbol{f}_{\mathrm{nom}}$, $\boldsymbol{h}_{\mathrm{nom}}$, and $\boldsymbol{\phi}$ are evaluated $2N$ times per time step. 
    The updates for $k_t$ and the sufficient statistics $\boldsymbol{\eta}_t$ involve matrix inversions, yielding a complexity of $\mathcal{O}(M^3)$ for $M$ basis functions. 
    Thus, the total cost per time step for Algorithm\,\ref{alg:online_RBPF} scales as $\mathcal{O}(N M^3)$.
    
    
	\section{Experimental Results}\label{sec:results}
    
	In this section, we test Algorithm\,\ref{alg:online_RBPF} and the prediction performance of the online-learned model in the real-world soft robot, depicted in Figure\,\ref{fig:problem_setting}. 
    The robot, the employed constant-curvature model, and the data are the same as in \cite{Mehl.2024}. 

    After a brief description of our initialization in Section\,\ref{sec:results_initialization}, we present and compare our results with \cite{Mehl.2024} in Section\,\ref{sec:results_estimation}. 
    We stress that, while \cite{Mehl.2024} presents an \ac{ukf} for state estimation, we additionally learn a state-dependent bending stiffness model. 
    Lastly, in Section\,\ref{sec:results_prediction}, we test the accuracy of the learned model by quantitatively analyzing multi-step prediction errors.

    As illustrated in Figure\,\ref{fig:problem_setting}, the driving system inputs $\boldsymbol{u}$ are pneumatic pressures, and the outputs $\boldsymbol{y}$ are base reaction forces and torques. 
    For validation purposes only, we have ground-truth robot states available, obtained through optical tracking. 
    The discretization time step of the data set is $\Delta t = 8 \, \mathrm{ms}$.

    \subsection{Initialization}\label{sec:results_initialization}
    To initialize our model, we set a non-informative \ac{gp} prior with initial hyperparameters $\ell_0= 1$, $\sigma^2_{f,0} = 100$, as well as $\mathcal{NIG}$ distribution parameters $\psi = 4$ and $\nu = 1$. 
    In the reduced-rank \ac{gp}, we use $M=40$ basis functions and set the domain as a hypercube with $L_i = 1$, $i=1,2,3$.  
    To comply with the latter, we apply a scaling operation to the \ac{gp} inputs $\boldsymbol{q}$, based on our knowledge of the rough workspace size. 
    The number of particles is set to $N = 500$, and the forgetting factor is chosen as $\gamma = 0$. 
    We point out that, while we run the marginalized \ac{pf} on $5\,\mathrm{s}$ of real-world data, a forgetting factor $\gamma > 0$ might be useful for longer time sequences to avoid accumulation of potential errors. 
     
    The nominal constant-curvature model $\boldsymbol{f}_{\mathrm{nom}}$ \& $\boldsymbol{h}_{\mathrm{nom}}$ is adopted from \cite{Mehl.2024}, including its offline identified model parameters. 
    The covariance matrices $\boldsymbol{\Sigma}_{\omega}$ and $\boldsymbol{\Sigma}_{e}$ are tuned heuristically. 
    For a fair comparison, we re-tuned the covariance matrices of the \ac{ukf} presented in \cite{Mehl.2024} heuristically, to perform well on the selected $5\,\mathrm{s}$ of real-world data. 
    The overall number of hyperparameter values to be tuned is $28$ for the marginalized \ac{pf} and $23$ for the \ac{ukf} \cite{Mehl.2024}, respectively. 

    Throughout heuristic tuning of the hyperparameters, we found that Algorithm\,\ref{alg:online_RBPF} is fairly insensitive to variations in the $\mathcal{NIG}$ hyperparameters $\nu$ and $\psi$.
    In contrast, the performance was influenced by the initialization of the \ac{gp} hyperparameters $\ell_0$ and $\sigma^2_{f,0}$, especially if no hyperparameter learning is used. 
    Moreover, the number of required basis functions $M$ rises with growing input domain, \ie with the size of $L_i$ and particularly with the number of \ac{gp} features \cite{RiutortMayol.2023}.
            
    \subsection{State Estimation and Model Learning}\label{sec:results_estimation}

\begin{figure}[tb] 
		\centering
		\includegraphics[width=0.99\linewidth]{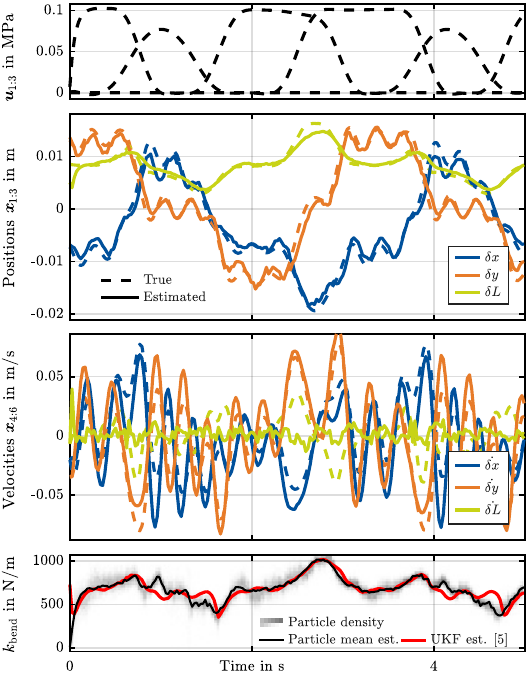}
		\caption{True and estimated hidden states of the soft robot, which is driven by the pneumatic control inputs. Simultaneously with state estimation, the marginalized \ac{pf} learns a nonlinear bending stiffness model $k_{\mathrm{bend}} (\boldsymbol{q})$.}
		\label{fig:state_trajectories}
	\end{figure}

    To evaluate the state estimation performance, we run Algorithm\,\ref{alg:online_RBPF} for $5\,\mathrm{s}$ on highly dynamic real-world soft robot data (see Figure\,\ref{fig:state_trajectories}). 
    As shown in the middle plots, the marginalized \ac{pf} accurately estimates the robot's current pose and velocity. 
    To compare, the average \ac{rmse} of the position and velocity states, as well as the overall \ac{nmse}, are given in Table\,\ref{tab:hyperparameters}. 
    As shown, the marginalized \ac{pf} yields comparable state estimation accuracy to the \ac{ukf} proposed in \cite{Mehl.2024}. 
    The results suggest improved estimation performance with the hyperparameter learning scheme. 
    Still, we emphasize that these differences may be affected by heuristic tuning or randomness. 

    For the bending stiffness, no ground-truth data is available. 
    In fact, the bending stiffness estimates have no strong physical meaning, as they serve as auxiliary slack variables to account for modeling errors. 
    Still, the bending stiffness $k_{\mathrm{bend}}(\boldsymbol{x})$ obeys a state-dependent relationship \cite{Mehl.2024}, which can be learned to improve the predictive accuracy of the overall soft robot model. 
    In this regard, the current bending stiffness estimates along the robot's trajectory are depicted in the bottom plot of Figure\,\ref{fig:state_trajectories}. 
    As shown, the particle approximation is consistent with the \ac{ukf} estimates from \cite{Mehl.2024}.
    
    \begin{table}[bt]
        \centering
        \caption{Comparison of the state estimation accuracy. Results for the marginalized \acp{pf} are averaged over $10$ Monte Carlo runs.}
        \label{tab:hyperparameters}
        \begin{tabular}{l | c | c | c c c}
            \toprule
            \textbf{Method} &\rot{\textbf{Bending} \\ \textbf{stiffness} \\  \textbf{model} \\ \textbf{learning}} & \rot{\textbf{Hyper-} \\ \textbf{parameter} \\ \textbf{learning}}& \rot{\textbf{Average} \\ \textbf{\acs{rmse} of} \\ \textbf{positions} } & \rot{\textbf{Average} \\ \textbf{\acs{rmse} of} \\ \textbf{velocities} } & \rot{\textbf{NMSE}} \\
            \midrule
            Marg. \ac{pf} & \cmark & \cmark &$1.28\,\mathrm{mm}$ & $18.3\,\mathrm{mm}/\mathrm{s}$ & $0.51$ \\
            Marg. \ac{pf} & \cmark & \xmark &$1.95\,\mathrm{mm}$ & $21.0\,\mathrm{mm}/\mathrm{s}$ & $0.64$ \\
            \ac{ukf} \cite{Mehl.2024} & \xmark &  -  & $1.33\,\mathrm{mm}$ & $40.6\,\mathrm{mm}/\mathrm{s}$ & $0.87$ \\
            \bottomrule
        \end{tabular}
        \vspace{-2mm}
    \end{table}

\begin{figure}[tb] 
		\centering
		\includegraphics[width=1\linewidth]{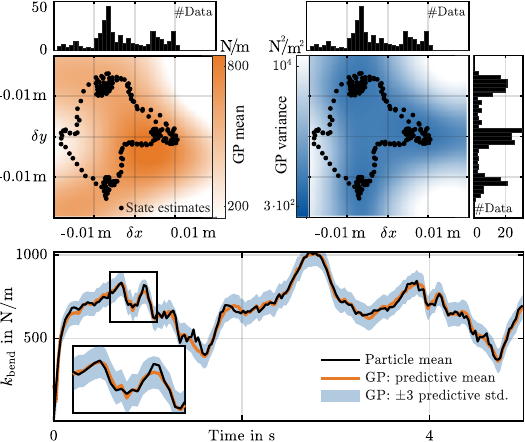}
		\caption{Mean and variance of the learned nonlinear bending stiffness \ac{gp} model (top, evaluated for $\delta L = 0.008\,\mathrm{m}$) and comparison of posterior \ac{gp} predictions with the particle-based estimates of the bending stiffness (bottom). The particle-based estimates and the \ac{gp} predictions are consistent with each other.}
		\label{fig:learned_stiffness}
	\end{figure}

    Figure\,\ref{fig:learned_stiffness} illustrates the learned bending stiffness \ac{gp} model at time $T=5\,\mathrm{s}$. 
    In the top plots, the mean and variance of the reduced-rank \ac{gp} are visualized over the generalized coordinates $\delta x$ and $\delta y$, while $\delta L$ is fixed at $0.008\,\mathrm{m}$. 
    As expected, regions with more state data points show a pronounced mean estimate and a low variance. 
    We stress that the state data points indicated by the histograms and the black dots are the state \textit{estimates} obtained by particle filtering. 
    This demonstrates the difficulty of learning a bending stiffness model with unknown regression inputs and outputs. 
    The bottom plot of Figure\,\ref{fig:learned_stiffness} compares the particle-based stiffness estimates $k_t$, obtained by sampling \eqref{eq:online_predictive_distribution}, with the learned bending stiffness \ac{gp} model at time $T=5\,\mathrm{s}$. 
    It can be seen that both particle estimates and \ac{gp} predictions are consistent.

    \subsection{Multi-Step Prediction}\label{sec:results_prediction}

    To assess the accuracy of the learned model, we perform multi-step forward predictions and compute the \ac{nmse} with respect to the real-world state measurements in a separate $5\,\mathrm{s}$-test trajectory. 
    In particular, we compare 
    \begin{enumerate}[label=\textit{(\roman*)}]
        \item the learned model $\boldsymbol{f}_{\mathrm{nom}}( \boldsymbol{x}_{t},  \boldsymbol{u}_{t}, \hat{k}_{\mathrm{bend}}( \boldsymbol{x}_{t}))$, where $\hat{k}_{\mathrm{bend}}( \boldsymbol{x}_{t})$ is the current \ac{gp} posterior mean \eqref{eq:rr_gp_pred},
        \item the nominal model $\boldsymbol{f}_{\mathrm{nom}}( \boldsymbol{x}_{t},  \boldsymbol{u}_{t}, \hat{k}_{\mathrm{bend,UKF}})$, where $\hat{k}_{\mathrm{bend,UKF}}$ is fixed at the current \ac{ukf} bending stiffness estimate, and
        \item the nominal model $\boldsymbol{f}_{\mathrm{nom}}( \boldsymbol{x}_{t},  \boldsymbol{u}_{t}, \bar{k}_{\mathrm{bend}})$, where $\bar{k}_{\mathrm{bend}} = 702.4\,\mathrm{N}/\mathrm{m}$ is fixed at the average estimated bending stiffness value.
    \end{enumerate}

    Table\,\ref{tab:pre_test_unique_label} summarizes the \ac{nmse} values for different integrator step sizes $\Delta t$ and prediction horizons $h$. 
    For instance, we evaluate $100$ prediction scenarios for $\Delta t = 5\,\mathrm{ms}$ and $h=20$, which we illustrate in terms of the state error evolution in a plot on the right-hand side of Table\,\ref{tab:pre_test_unique_label}. 
    As shown, all models exhibit similar prediction accuracy for small step sizes and short prediction horizons. 
    However, the nominal model \textit{(iii)} exhibits significant prediction errors if the integrator time step or the horizon length is increased. 
    This can be attributed to strong modeling assumptions and, thus, to a too simple model for representing the complex real-world behavior of the soft robot. 
    In comparison, we observe that both the nominal model \textit{(ii)} that incorporates \ac{ukf} estimates and the \ac{gp}-augmented overall model \textit{(i)} provide accurate multi-step predictions. 
    When considering larger step sizes and prediction horizons, the \ac{gp}-augmented model \textit{(i)} outperforms the nominal model \textit{(iii)} by a factor of two or more, and outperforms the nominal model with \ac{ukf} estimates \textit{(ii)} by approx. $10\,\%$.

	\section{Conclusion and Outlook}\label{sec:conclusion}
    
	In this paper, we show how to estimate a soft robot's current pose while---in contrast to existing work---simultaneously learning a bending stiffness model online. 
    This is enabled by a recently proposed gray-box system identification tool \cite{Volkmann.2025}, which we validate in a complex, real-world robotic system. 
    The learned nonlinear stiffness model improves overall model quality compared to a nominal model and thereby allows for accurate multi-step forward predictions. 
    Methodologically, we build on prior work regarding simultaneous state estimation and model learning \cite{Berntorp.2021,Volkmann.2025} and employ a particle filter, marginalized over the model parameters.

    We acknowledge that, in its current implementation, the estimation scheme is not suitable for real-time applications, but speed-up is possible. 
    Moreover, further work is required to systematically tune the algorithmic parameters. 
    
    Nonetheless, taking a step back, the results demonstrate that recent online inference and learning tools enable accurate, adaptive modeling, paving the way for predictive model-based control strategies.

	\begin{table*}[tb]
		\caption{Comparison of state prediction \ac{nmse} for the nominal and \ac{gp}-enhanced constant-curvature soft robot across pred. horizons and discretization widths. Mean \& standard deviation across states are given. The plot illustrates one prediction scenario.}
		\label{tab:pre_test_unique_label} 
        
        {\fontsize{7pt}{7pt}\selectfont
		\begin{tikzpicture}
			\node[inner sep=0pt] (tbl) {
				\resizebox{0.60\textwidth}{!}{%
\begin{tabular}{c | l | cccccc}
    \toprule
    $h$ & \textbf{Model} & $\Delta t = {1.0}\,\mathrm{ms}$ & $\Delta t = {2.5}\,\mathrm{ms}$ & $\Delta t = {5.0}\,\mathrm{ms}$ & $\Delta t = {7.5}\,\mathrm{ms}$ & $\Delta t = {10.0}\,\mathrm{ms}$ \\
    \midrule
    {5}     & Nom. \& \acs{gp} & \cellcolor[rgb]{0.66,0.92,0.20} 0.058 \scriptsize(0.052) & \cellcolor[rgb]{1.00,0.98,0.20} 0.103 \scriptsize(0.091) & \cellcolor[rgb]{0.99,1.00,0.20} 0.098 \scriptsize(0.082) & \cellcolor[rgb]{0.91,0.98,0.20} 0.089 \scriptsize(0.064) & \cellcolor[rgb]{0.96,0.64,0.20} 0.145 \scriptsize(0.126) \\
 & Nom. \& \acs{ukf} est. & \cellcolor[rgb]{0.67,0.92,0.20} 0.059 \scriptsize(0.054) & \cellcolor[rgb]{0.99,0.95,0.20} 0.106 \scriptsize(0.094) & \cellcolor[rgb]{1.00,0.96,0.20} 0.105 \scriptsize(0.083) & \cellcolor[rgb]{1.00,1.00,0.20} 0.099 \scriptsize(0.069) & \cellcolor[rgb]{0.94,0.50,0.20} 0.162 \scriptsize(0.136) \\
 & Nominal & \cellcolor[rgb]{0.68,0.92,0.20} 0.060 \scriptsize(0.050) & \cellcolor[rgb]{0.99,0.90,0.20} 0.113 \scriptsize(0.092) & \cellcolor[rgb]{0.97,0.78,0.20} 0.128 \scriptsize(0.099) & \cellcolor[rgb]{0.96,0.70,0.20} 0.137 \scriptsize(0.102) & \cellcolor[rgb]{0.90,0.20,0.20} 0.211 \scriptsize(0.183) \\
    \midrule
    {10}     & Nom. \& \acs{gp} & \cellcolor[rgb]{0.87,0.97,0.20} 0.084 \scriptsize(0.075) & \cellcolor[rgb]{0.95,0.99,0.20} 0.094 \scriptsize(0.080) & \cellcolor[rgb]{0.85,0.96,0.20} 0.082 \scriptsize(0.059) & \cellcolor[rgb]{0.85,0.96,0.20} 0.081 \scriptsize(0.052) & \cellcolor[rgb]{0.98,0.84,0.20} 0.119 \scriptsize(0.091) \\
 & Nom. \& \acs{ukf} est. & \cellcolor[rgb]{0.89,0.97,0.20} 0.087 \scriptsize(0.077) & \cellcolor[rgb]{1.00,1.00,0.20} 0.101 \scriptsize(0.081) & \cellcolor[rgb]{0.95,0.99,0.20} 0.093 \scriptsize(0.062) & \cellcolor[rgb]{0.97,0.99,0.20} 0.096 \scriptsize(0.057) & \cellcolor[rgb]{0.96,0.72,0.20} 0.135 \scriptsize(0.097) \\
 & Nominal & \cellcolor[rgb]{0.92,0.98,0.20} 0.091 \scriptsize(0.075) & \cellcolor[rgb]{0.98,0.83,0.20} 0.122 \scriptsize(0.095) & \cellcolor[rgb]{0.96,0.67,0.20} 0.141 \scriptsize(0.111) & \cellcolor[rgb]{0.92,0.40,0.20} 0.175 \scriptsize(0.158) & \cellcolor[rgb]{0.90,0.20,0.20} 0.248 \scriptsize(0.286) \\
    \midrule
    {20}     & Nom. \& \acs{gp} & \cellcolor[rgb]{0.97,0.99,0.20} 0.096 \scriptsize(0.084) & \cellcolor[rgb]{0.84,0.96,0.20} 0.080 \scriptsize(0.057) & \cellcolor[rgb]{0.86,0.97,0.20} 0.083 \scriptsize(0.053) & \cellcolor[rgb]{0.91,0.98,0.20} 0.089 \scriptsize(0.056) & \cellcolor[rgb]{0.90,0.20,0.20} 0.305 \scriptsize(0.892) \\
 & Nom. \& \acs{ukf} est. & \cellcolor[rgb]{1.00,0.99,0.20} 0.101 \scriptsize(0.085) & \cellcolor[rgb]{0.93,0.98,0.20} 0.091 \scriptsize(0.061) & \cellcolor[rgb]{0.95,0.99,0.20} \tikzmarknode{tabMark}{ 0.094 \scriptsize(0.052)} & \cellcolor[rgb]{0.98,0.99,0.20} 0.097 \scriptsize(0.048) & \cellcolor[rgb]{0.97,0.76,0.20} 0.130 \scriptsize(0.071) \\
 & Nominal & \cellcolor[rgb]{0.99,0.88,0.20} 0.115 \scriptsize(0.091) & \cellcolor[rgb]{0.96,0.71,0.20} 0.136 \scriptsize(0.106) & \cellcolor[rgb]{0.90,0.23,0.20} \tikzmarknode{tabMark}{ 0.196 \scriptsize(0.202)} & \cellcolor[rgb]{0.90,0.20,0.20} 0.218 \scriptsize(0.238) & \cellcolor[rgb]{0.90,0.20,0.20} \textbf{$>1$} \\
    \bottomrule
\end{tabular}

				}
			};
			
			\node[inner sep=0pt, right=0.2cm of tbl] (fig) {
				\includegraphics[width=0.4\textwidth]{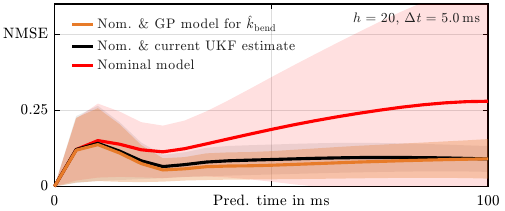}
			};
			
			\coordinate (target) at ([xshift=1.91cm, yshift=0.42cm]tbl.south); 
			
			\draw[black, thick] ($(target |- target) + (-1.62cm, 0.32cm)$) rectangle ($(target) + (0cm, -0.32cm)$);

			\draw[black!80, thin] ($(fig.north west) + (0.8cm, -0.07cm)$) -- ($(target) + ( 0.0cm, +0.32cm)$); 
			\draw[black!80, thin] ($(fig.south west) + (0.8cm, +0.34cm)$) -- ($(target) + ( -0.0cm, -0.32cm)$); 
			
		\end{tikzpicture}}
	\end{table*}

	
    
	{\printbibliography}

\end{document}